\documentclass[a4paper,11pt]{article}
\usepackage{pos}
\usepackage{ragged2e}
\usepackage{subcaption}

\title{Higgs boson decays into gauge bosons: theoretical predictions in RNS}

\author*[a]{Edilson A. Reyes R.}

\affiliation[a]{Universidad de Pamplona,\\
  Departamento de Física, Pamplona, Colombia}

\emailAdd{edilson.reyes@unipamplona.edu.co}

\abstract{
We discuss recent results on the loop-induced decays $h\to Z\gamma$, $h\to\gamma\gamma$, and $h\to gg$ in the Radiatively Natural
Supersymmetry (RNS) framework. These processes are sensitive to contributions from particles beyond the Standard Model and provide complementary probes of supersymmetric effects. We evaluate the decay amplitudes and partial widths for parameter points consistent with a Higgs boson mass near the measured value of 125 GeV, and examine their dependence on the model parameters. Across the
parameter space, the predicted $h\to Z\gamma$ width can reach about 7.5 keV, while the $h\to\gamma\gamma$ and $h\to gg$ widths remain close to their Standard Model predictions. We compare the $h\to Z\gamma$ predictions with the available ATLAS and CMS measurements and discuss the prospects for distinguishing RNS from the Standard Model with improved experimental precision.
}

\FullConference{43rd International Conference on High Energy Physics (ICHEP 2026)\\
30 July  to 5 August , 2026\\
Natal, Brazil\\}

\begin{document}
\maketitle

\section{Introduction}
\begin{figure}[t]
    \centering
    \includegraphics[width=0.7\textwidth]{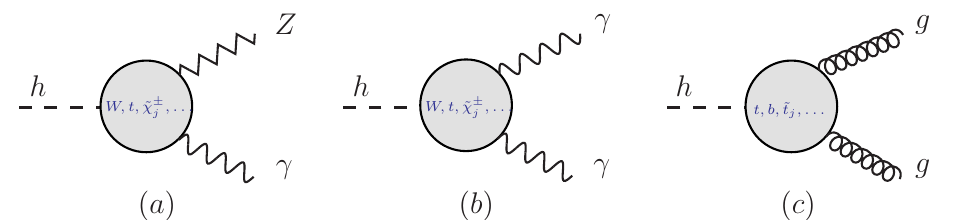}
    \caption{ 
        \justifying \small{Schematic representation of the loop-induced Higgs boson decays into pairs of gauge bosons.}
       }
    \label{fig:HV1V2}
\end{figure}
The discovery of the Higgs boson at the LHC completed the particle content of the Standard Model (SM) and opened a new era of precision studies of the electroweak and Higgs observables. Comparing precise theoretical predictions with current and future measurements of the Higgs boson properties provides an important opportunity to look for effects of physics beyond the SM. In this context, the loop-induced Higgs boson decays (a) $h\to Z\gamma$, (b) $h\to\gamma\gamma$, and (c) $h\to gg$, schematically illustrated in Figure~\ref{fig:HV1V2}, are particularly interesting because they do not occur at tree level and can therefore be sensitive to new particles and interactions that modify the corresponding amplitudes. Among these processes, the rare decay $h\to Z\gamma$ has recently received particular attention following the first evidence reported by the ATLAS and CMS collaborations in 2023~\cite{ATLASCMS2023}. More recently, ATLAS has updated its search using Run 2 and Run 3 data, reporting a combined signal strength of $\mu=1.3^{+0.6}_{-0.5}$~\cite{ATLAS2026}, while CMS has reported $\mu=1.10^{+0.52}_{-0.61}$~\cite{CMS2026}. Although the central values remain above the SM prediction, both measurements are consistent with the SM within their large uncertainties, which are still of the order of 40\%. Future collider experiments are expected to substantially improve the measurement precision, potentially reaching below 10\%~\cite{FCC2019}. As discussed in~\cite{Reyes2026}, this level of precision could allow deviations predicted by supersymmetric scenarios to be distinguished from the SM expectation at the $3\sigma$ level or beyond, thereby making the $h\to Z\gamma$ decay a promising probe of new physics. \\ The theoretical prediction of the $h\to Z\gamma$ partial decay width ($\Gamma_{Z\gamma}$) in the SM has been studied extensively, including the leading-order result and higher-order QCD and electroweak corrections~\cite{Bergstrom1985,Djouadi1998,Gehrmann2015,Chen2024,Sang2024}. The dominant contribution come from the $W$ boson one-loop correction, while loops with heavy charged fermions also play an important role. This decay has also been investigated in several extensions of the SM, where additional charged particles can contribute to the loop amplitudes. In models including supersymmetry (SUSY)~\cite{Cao2013,Hammad2015,Liu2020,Israr2025}, these contributions can arise from particles such as charginos and scalar superpartners, potentially modifying the value of $\Gamma_{Z\gamma}$ with respect to the SM prediction. Important modifications to $\Gamma_{Z\gamma}$ can also be obtained including non-SUSY interactions and possible sources of CP violation~\cite{HernandezJuarez2025,HernandezJuarez2026}. \\
In this article, we review the predictions of $\Gamma_{Z\gamma}$ in the Radiative Natural Supersymmetry scenario~\cite{Baer2012,Baer2013}, a well-motivated framework realized within the Minimal Supersymmetric Standard Model (MSSM) that enhance the $h\to Z\gamma$ decay width relative to the SM prediction, while accommodates a Higgs boson mass compatible with the LHC measurement, $M_h\sim125$ GeV, allowing a relatively low electroweak fine-tuning parameter~\cite{Reyes2025}. Building on this analysis, we subsequently show the correlated predictions for the loop-induced decays within RNS. This complementary study allows us to investigate whether parameter regions that enhance $\Gamma_{Z\gamma}$ remain compatible with the experimental constraints from the other two loop-induced channels, $h\to\gamma\gamma$ and $h\to gg$, whose measured rates are in good agreement with the SM predictions and are subject to relatively small uncertainties.

\section{Parameter Space in RNS}

In the RNS scenario, the MSSM parameters are specified through a set of unified boundary conditions at the grand unification scale, $\Lambda_{\mathrm{GUT}}=1.5\times10^{16}$ GeV, and subsequently evolved to the SUSY scale using the renormalization group equations. The independent input parameters are the universal scalar mass $m_0$, the universal gaugino mass $m_{1/2}$, the universal trilinear coupling $A_0$, the ratio of the vacuum expectation values of the two Higgs doublets, $\tan\beta=v_u/v_d$, the Higgsino mass parameter $\mu$, and the mass of the CP-odd Higgs boson, $m_A$. In this analysis, we consider the region of parameter space that maximizes the partial decay width $\Gamma_{Z\gamma}$ while reproducing the observed Higgs boson mass within $M_h=125\pm3$ GeV and maintaining a moderately large electroweak fine-tuning parameter, $\Delta_{\mathrm{EW}}\sim100$. The considered $M_h$ range takes into account the estimated theoretical uncertainties in MSSM Higgs mass calculations, arising from missing higher-order radiative corrections and parametric uncertainties~\cite{Slavich2021}. The corresponding parameter ranges are shown in Table \ref{tab:RNSparameters}.
\begin{table}[t]
\centering
\renewcommand{\arraystretch}{1.15}
\begin{tabular*}{\textwidth}{@{\extracolsep{\fill}} |ccccccc|}
\hline
$m_0$ [TeV] & $m_{1/2}$ [TeV] & $|A_0/m_0|$ & $\tan\beta$ &
$m_A$ [TeV] & $\mu$ [GeV] & $M_h$ [GeV] \\
\hline
$2-8$ & $0.5-2.5$ & $1.75$ & $8-55$ &
$0.6-1.5$ & $50-350$ & $125\pm3$ \\
\hline
\end{tabular*}
\caption{Parameter ranges considered in the RNS analysis.}
\label{tab:RNSparameters}
\end{table} \\ 
The evolution of the MSSM parameters and the calculation of the supersymmetric mass spectrum are performed using ISASUGRA, included in the ISAJET package~\cite{Baer:2003}. The RGEs are integrated from $\Lambda_{\mathrm{GUT}}$ down to the supersymmetry matching scale, defined as the geometric mean of the two stop masses, $\Lambda_{\mathrm{SUSY}}=\sqrt{m_{\tilde t_1}m_{\tilde t_2}}$. The running parameters are treated within the $\overline{\mathrm{DR}}$ renormalization scheme and converted into physical pole masses by including the relevant radiative corrections. An important feature of RNS is the requirement of a relatively small Higgsino mass parameter, $\mu$, which is closely related to electroweak naturalness. In particular, low values of $\mu$ reduce the electroweak fine-tuning, while experimental constraints from LEP2 and the LHC require $\mu\gtrsim100$ GeV, approximately corresponding to the lower limit on the lightest chargino mass. In the limit $M_1,M_2\gg\mu$, where $M_1$ and $M_2$ denote the bino and wino mass parameters, respectively, the lightest neutralino is predominantly Higgsino-like, with a mass approximately determined by $\mu$. Consequently, RNS predicts relatively light Higgsino states, which are relevant both for collider phenomenology and dark matter searches. However, this feature also introduces a potential tension with direct detection constraints. In particular, the stringent limits from the LUX-ZEPLIN~\cite{LZ2025} experiment place significant pressure on natural supersymmetric scenarios with Higgsino-like neutralino dark matter. Recent analyses have investigated whether cancellations in the neutralino-nucleon scattering amplitudes, commonly referred to as blind spots, can alleviate this tension and preserve viable regions of natural supersymmetry~\cite{Baer2026LZ}. Thus, although small values of $\mu$ are favored from the perspective of electroweak naturalness, their compatibility with dark matter constraints represents an additional consideration in assessing the phenomenological viability of RNS. Additionally, all decays of the Higgs boson into supersymmetric particles are kinematically forbidden, since the lightest neutralino mass satisfies $m_{\tilde{\chi}_1^0}>M_h/2$, while all other superpartners are considerably heavier. Consequently, the Higgs boson can decay only into SM particles, and its total decay width receives no direct contribution from SUSY final states. This makes loop-induced processes particularly relevant, since potential SUSY effects can enter only through virtual corrections in the decay amplitudes. 

\section{Numerical Analysis}

\begin{table}[]
\centering
{\small
\begin{tabular}{|c|c|c|c|}
\hline
\textbf{Contribution} & \textbf{$h\to Z\gamma$} & \textbf{$h\to \gamma\gamma$} & \textbf{$h\to gg$} \\ \hline
\multicolumn{4}{|c|}{\texttt{ Main SM-like contributions}}\\ \hline
$W$-boson   & $5.1616$  &  $-8.3452$   &   --   \\
top   &  $-0.08182$ &  $1.8393$  &    $1.3766$   \\ 
bottom &  $-0.0036+0.0021\,i$ &  $-0.0248+0.0318\,i$    &  $-0.0732+0.0945\,i$   \\ \hline
\multicolumn{4}{|c|}{\texttt{ Main SUSY contributions}}\\ \hline
chargino $\tilde{\chi}_1$ & $0.0017$ &  $-0.0225$   &  --    \\ 
chargino $\tilde{\chi}_2$ &  $-0.0021$ &  $0.0109$    & -- \\ 
stop $\tilde t_1$ & $0.0011$ &   $-0.0180$  &  $- 0.0131$   \\ 
stop $\tilde t_2$ &   $0.00027$   &   $1.62\times10^{-3}$   &  $1.391\times10^{-3}$  \\  
sbottom $\tilde b_1$ & $2.85\times 10^{-5}$ &  $-3.36\times10^{-5}$    &   $-1.18\times10^{-4}$   \\
stau $\tilde \tau_1$& $-7.76\times 10^{-6}$ &  $1.87\times10^{-4}$   &  --  \\ \hline
\multicolumn{4}{|c|}{\texttt{Total prediction}}\\ \hline
$\Gamma_{\rm }^{\rm RNS}$         &  $6.1878~\text{keV}$    &  $9.4589~\text{keV}$    &   $0.2924~\text{MeV}$  \\ \hline
\end{tabular}}
\caption{\justifying \small{Individual contributions to the amplitudes of the loop-induced Higgs boson decays for a representative benchmark point where $M_h\approx 125$ GeV. The quantities listed correspond to the main contributions entering the one-loop form factors, together with the resulting partial decay widths.}}
\label{tab:hAA}
\end{table} 
We begin by presenting the dominant one-loop contributions to the loop-induced Higgs decay amplitudes, summarized in Table~\ref{tab:hAA}, for the benchmark point which yields a Higgs boson mass of approximately $125$ GeV, consistent with the LHC measurement. The corresponding parameter values are $m_0=4$ TeV, $A_0=-7$ TeV, $\mu=100$ GeV, $m_{1/2}=1.5$ TeV, $m_A=600$ GeV, and $\tan\beta=30$. In the $h\to Z\gamma$ and $h\to\gamma\gamma$ channels, the $W$-boson loop provides the dominant contribution, while the top-quark loop is subdominant. For $h\to gg$, the top-quark loop dominates, with a smaller contribution from the bottom-quark loop. The SUSY contributions are subdominant in all three channels. In particular, the total SUSY amplitude amounts to approximately $0.02\%$, $0.43\%$, and $0.91\%$ of the corresponding SM-like amplitude for $h\to Z\gamma$, $h\to\gamma\gamma$, and $h\to gg$, respectively. In the electroweak channels, the lightest chargino provides the largest SUSY contribution, whereas in the gluonic channel, the dominant SUSY contribution comes from the lightest stop. The resulting partial widths for this benchmark are $\Gamma^{\rm RNS}_{Z\gamma}=6.1878$~keV, $\Gamma^{\rm RNS}_{\gamma\gamma}=9.4589$~keV, and $\Gamma^{\rm RNS}_{gg}=0.2924$~MeV. \\ We now present the main numerical results for the loop-induced Higgs boson decays within the complete RNS scenario. We focus on the parameter region that maximizes $\Gamma_{Z\gamma}$, while also examining the compatibility of the resulting predictions with the other loop-induced channels. 
\begin{figure*}[]
  \centering  
  \begin{subfigure}[c]{0.315\textwidth}
    \centering    \includegraphics[width=\linewidth]{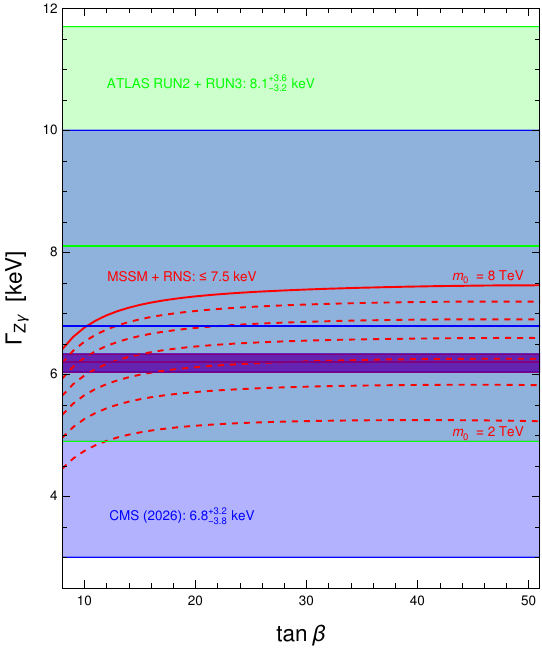}
    \caption{ }
    \label{fig:plotGhZgPoS}
  \end{subfigure} \hspace{0.2cm}
   \begin{subfigure}[c]{0.31\textwidth}
    \centering    \includegraphics[width=\linewidth]{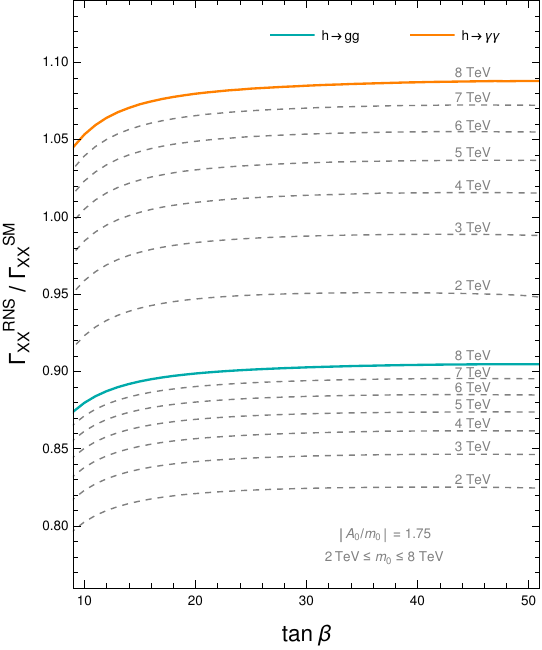}
    \caption{ }
    \label{fig:plot_tb_PoS}
  \end{subfigure} \hspace{0.3cm}
  \begin{subfigure}[c]{0.31\textwidth}
    \centering    \includegraphics[width=\linewidth]{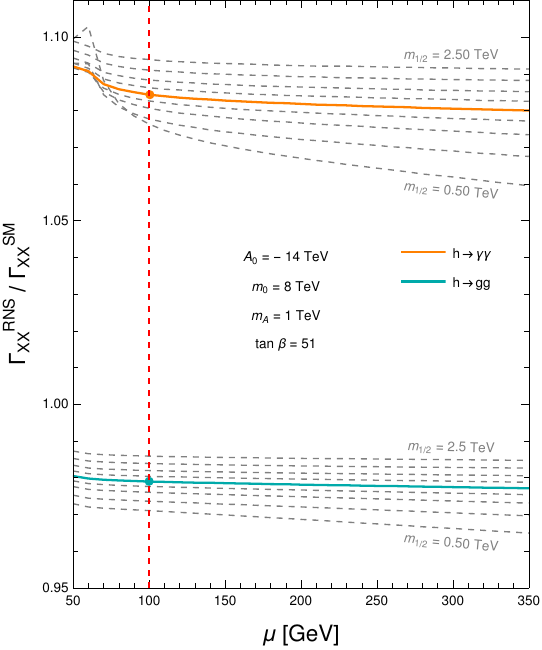}
        \caption{ }
        \label{fig:plotGG_PoS}
  \end{subfigure}
   \caption{  \justifying \small{Properties of the loop-induced Higgs boson decays in the RNS scenario. Panel~\textbf{(a)}: shows the $h \to Z\gamma$ decay width predictions as a function of $\tan\beta$ for different values of $m_0$ in the RNS scenario. Panel~\textbf{(b)}: shows the ratio of the corresponding decay width in RNS to the SM prediction as a function of $\tan\beta$ for different values of the unified scalar mass $m_0$. Panel~\textbf{(c)} shows the dependence of the normalized Higgs boson decay widths on the higgsino mass parameter $\mu$ for different values of the unified gaugino mass~$m_{1/2}$. }}
    \label{fig:GG_plots}
\end{figure*}
Figure~\ref{fig:plotGhZgPoS} shows the dependence of $\Gamma_{Z\gamma}$ on $\tan\beta$ for different values of $m_0$. The RNS predictions are represented with red lines, reaches a maximum of approximately $7.5$ keV for $m_0=8$ TeV and $\tan\beta=51$, corresponding to an enhancement of about $20\%$ regarding the central SM value. The purple band represents the SM prediction with its theoretical uncertainty, $\Gamma^{\rm SM}_{Z\gamma}=6.19\pm0.15$ keV. The green band corresponds to the combined Run2 + Run3 ATLAS measurement, $\Gamma^{\rm ATLAS}_{Z\gamma}=8.1^{+3.6}_{-3.2}$ keV, which is compatible with both the SM and RNS predictions. Finally, the blue band corresponds to the latest CMS measurement, $\Gamma^{\rm CMS}_{Z\gamma}=6.8^{+3.2}_{-3.8}$ keV, which has a central value closer to the SM expectation, but its large uncertainty also allows for the RNS enhancement. Assuming that the current central values remain unchanged, a $3\sigma$ discrimination between the SM and the ATLAS measurement would require an experimental uncertainty below approximately $0.6$ keV, corresponding to a relative precision of about $8\%$. For the CMS measurement, whose central value is closer to the SM prediction, a substantially higher precision would be required, with an uncertainty below approximately $0.2$ keV, or about $3\%$. These results highlight the importance of improving the precision of $h\to Z\gamma$ measurements at future collider experiments. \\ The correlated predictions for the other two channels are shown in Figures~\ref{fig:plot_tb_PoS} and~\ref{fig:plotGG_PoS}. Figure~\ref{fig:plot_tb_PoS} displays the ratios $\Gamma^{\rm RNS}_{XX}/\Gamma^{\rm SM}_{XX}$, with $X=\gamma,g$, as functions of $\tan\beta$ for different values of $m_0$. The diphoton channel exhibits a mild enhancement, reaching approximately $8\%$ for $m_0=8$ TeV, whereas the gluonic channel is systematically suppressed, with deviations of about $10\%-20\%$ relative to the SM prediction. The parameter region that maximizes $\Gamma_{Z\gamma}$ predicts deviations of approximately $5\%$ and $10\%$ in the diphoton and gluonic channels, respectively. These results show that the enhancement of $h\to Z\gamma$ is accompanied by correlated effects in the other loop-induced decays. Figure~\ref{fig:plotGG_PoS} shows the dependence of the same ratios on the Higgsino mass parameter $\mu$. Both decay widths exhibit a weak dependence on $\mu$ in the phenomenologically relevant region $100\leq\mu\leq350$ GeV. For the benchmark configuration that maximizes $\Gamma_{Z\gamma}$, the variation of $\Gamma_{\gamma\gamma}$ is below $0.4\%$, while that of $\Gamma_{gg}$ is below $0.2\%$. Variations of $m_{1/2}$ over the considered range produce only moderate changes, indicating that the predicted deviations are stable against variations of these parameters. \\ Overall, the results show that the loop-induced Higgs decays considered here provide complementary probes of RNS, with the largest deviations from the SM occurring in $h\to Z\gamma$ and offering a target for future precision measurements.


\begin{thebibliography}{99}
{\footnotesize
\bibitem{ATLASCMS2023}
G. Aad et al. (ATLAS and CMS Collaborations),
Phys. Rev. Lett. \textbf{132}, 021803 (2024), arXiv:2309.03501 [hep-ex]. 

\bibitem{ATLAS2026}
G. Aad et al. (ATLAS Collaboration),
Phys. Lett. B \textbf{876}, 140313 (2026), arXiv:2507.12598 [hep-ex].

\bibitem{CMS2026}
A. Hayrapetyan et al. (CMS Collaboration),
arXiv:2609.04402 [hep-ex] (2026).

\bibitem{FCC2019}
A. Abada et al.,
Eur. Phys. J. Special Topics \textbf{228}, 261 (2019).

\bibitem{Reyes2026}
E.~A. Reyes R.,
Phys. Rev. D \textbf{114}, 055015 (2026),
arXiv:2604.22115 [hep-ph].

\bibitem{Bergstrom1985}
L. Bergstr\"om and G. Hulth,
Nucl. Phys. B \textbf{259}, 137 (1985);
Erratum, Nucl. Phys. B \textbf{276}, 744 (1986).

\bibitem{Djouadi1998}
A. Djouadi, V. Driesen and W. Hollik,
Eur. Phys. J. C \textbf{1}, 163 (1998),
arXiv:hep-ph/9701342.

\bibitem{Gehrmann2015}
T. Gehrmann, S. Guns and D. Kara,
JHEP \textbf{09}, 038 (2015),
arXiv:1505.00561 [hep-ph].

\bibitem{Chen2024}
Z.-Q. Chen, L.-B. Chen, C.-F. Qiao and R. Zhu,
Phys. Rev. D \textbf{110}, L051301 (2024),
arXiv:2404.11441 [hep-ph].

\bibitem{Sang2024}
W.-L. Sang, F. Feng and Y. Jia,
Phys. Rev. D \textbf{110}, L051302 (2024),
arXiv:2405.03464 [hep-ph].

\bibitem{Cao2013}
J. Cao, L. Wu, P. Wu and J.~M. Yang,
JHEP \textbf{09}, 043 (2013),
arXiv:1301.4641 [hep-ph].

\bibitem{Hammad2015}
A. Hammad, S. Khalil and S. Moretti,
Phys. Rev. D \textbf{92}, 095008 (2015),
arXiv:1503.05408 [hep-ph].

\bibitem{Liu2020}
C.-X. Liu, H.-B. Zhang, J.-L. Yang, S.-M. Zhao,
Y.-B. Liu and T.-F. Feng,
JHEP \textbf{04}, 002 (2020),
arXiv:2002.04370 [hep-ph].

\bibitem{Israr2025}
S. Israr and M. Rehman,
Eur. Phys. J. Plus \textbf{140}, 397 (2025),
arXiv:2407.01210 [hep-ph].

\bibitem{HernandezJuarez2025}
A. Hernandez-Juarez. R. Gaitan and R. Martinez. 
Phys. Rev. D \textbf{111}, 015001 (2025), arXiv:2405.03094 [hep-ph].

\bibitem{HernandezJuarez2026}
A. I. Hern\'andez-Ju\'arez, D. Watko, A. Fern\'andez-T\'ellez and G. Tavares-Velasco,
Phys. Rev. D \textbf{114}, 055036 (2026), arXiv:2608.17126 [hep-ph].

\bibitem{Baer2012}
H. Baer, V. Barger, P. Huang, A. Mustafayev and X. Tata,
Phys. Rev. Lett. \textbf{109}, 161802 (2012),
arXiv:1207.3343 [hep-ph].

\bibitem{Baer2013}
H. Baer, V. Barger, P. Huang, D. Mickelson,
A. Mustafayev and X. Tata,
Phys. Rev. D \textbf{87}, 115028 (2013),
arXiv:1212.2655 [hep-ph].

\bibitem{Reyes2025}
E.~A. Reyes R., C.~A. Lopez A., O.~R. Torrijo G. and D.~G. Melo P.,
Phys. Rev. D \textbf{112}, 095012 (2025),
arXiv:2507.09395 [hep-ph].

\bibitem{Slavich2021}
P. Slavich et al.,
Eur. Phys. J. C \textbf{81}, 450 (2021),
arXiv:2012.15629 [hep-ph].

\bibitem{Baer:2003}
H. Baer, F. E. Paige, S. D. Protopopescu, and X. Tata,
``ISAJET 7.69: A Monte Carlo Event Generator for $pp$, $\bar pp$, and $e^+e^-$ Reactions,''
arXiv:hep-ph/0312045.

\bibitem{LZ2025}
J. Aalbers et al. (LZ Collaboration),
Phys. Rev. Lett. \textbf{135}, 011802 (2025),
arXiv:2410.17036 [hep-ex].

\bibitem{Baer2026LZ}
H. Baer, V. Barger and D. Sengupta,
JCAP \textbf{09}, 137 (2026), arXiv:2607.05648 [hep-ph]. 

}
\end{thebibliography}
\end{document}